\documentclass[twocolumn,showpacs,epsfig,preprintnumbers]{revtex4} 
\usepackage{graphicx}
\usepackage{dcolumn}
\usepackage{bm}
\usepackage{amssymb}
\usepackage{amsmath}
\usepackage{epsfig}   
\usepackage[usenames]{color}

\definecolor{green}{cmyk}{1,0,1,0}

\def\beq{\begin{equation}}
\def\eeq{\end{equation}}
\def\bea{\begin{array}}
\def\eea{\end{array}}
\def\be{\begin{equation}}
\def\ee{\end{equation}}
\def\ba{\begin{eqnarray}}
\def\ea{\end{eqnarray}}

\def\to{\rightarrow}

\def\[{\left[}
\def\]{\right]}
\def\({\left(}
\def\){\right)}

\def\sm0{{\widetilde{m}_0}}

\def\U1em{{U(1)_{\rm em}}}
\def\to{\rightarrow}

\def\sq2{\sqrt{2}}

\def\ee{e^+e^-}

\def\End{\end{document}}

\def\fsl#1{\setbox0=\hbox{$#1$}                 
   \dimen0=\wd0                                 
   \setbox1=\hbox{/} \dimen1=\wd1               
   \ifdim\dimen0>\dimen1                        
      \rlap{\hbox to \dimen0{\hfil/\hfil}}      
      #1                                        
   \else                                        
      \rlap{\hbox to \dimen1{\hfil$#1$\hfil}}   
      /                                         
   \fi}

\begin{document}                                                              

\title{Masses of dark matter and neutrino  
from TeV scale spontaneous $U(1)_{B-L}$ breaking 
}%
\preprint{UT-HET 049, HGU-CAP 008, YITP-11-10}
\author{%
{\sc Shinya Kanemura\,$^1$, Osamu Seto\,$^2$, and Takashi Shimomura\,$^3$}
}
\affiliation{%
\vspace*{2mm} 
$^1$Department of Physics, University of Toyama, 3190 Gofuku, Toyama 930-8555, Japan\\ 
$^2$Department of Architecture and Building Engineering, Hokkai-Gakuen University, Sapporo 062-8605, Japan\\
$^3$Yukawa Institute for Theoretical Physics, Kyoto University, Kyoto 606-8502, Japan\\}

\begin{abstract}
We propose a simple testable 
model with  mass generation mechanisms for dark matter and neutrino 
based on the gauged $U(1)_{B-L}$ symmetry and an exact $Z_2$ parity. 
The $U(1)_{B-L}$ symmetry is spontaneously broken at the TeV scale,  
by which $Z_2$-odd right-handed neutrinos 
receive Majorana masses of the electroweak scale. 
The lightest one 
is a dark matter candidate, 
whose stability is guaranteed by the $Z_2$ parity.
Resulting lepton number violation is transmitted 
to the 
left-handed neutrinos $\nu_L^i$ via 
the loop-induced dimension-six operator.  
Consequently, the tiny masses of $\nu_L^i$ can be 
generated 
without excessive fine tuning.  
The observed dark matter abundance can be reproduced 
by  the pair annihilation via 
the s-channel scalar exchange 
due to mixing of neutral components of $\Phi$ and $S$, where 
$\Phi$ and $S$ respectively represent the Higgs doublet  
and the additional scalar singlet with the $B-L$ charge.
The model can be tested at collider experiments 
as well as flavor experiments  
through the discriminative predictions such as 
two light neutral Higgs bosons with large mixing, 
invisible decays of the Higgs bosons as well as the $B-L$ 
gauge boson, and lepton flavor violation. 

\pacs{\,   }
\end{abstract}

\maketitle

\setcounter{footnote}{0}
\renewcommand{\thefootnote}{\arabic{footnote}}


Physics of electroweak symmetry breaking (EWSB), 
which is responsible for generating 
masses of weak bosons as well as quarks and charged leptons,    
is the last unknown part in the standard model 
(SM) for elementary particles.  Its exploration 
is one of the top priorities 
at the Fermilab Tevatron and 
the CERN Large Hadron Collider (LHC). 

Despite of its great success, 
the SM cannot explain several phenomena established experimentally.  
First, previous neutrino oscillation experiments 
have clarified that neutrinos have tiny masses  
 (less than 1 eV), which are much lower than 
the electroweak scale, 100 GeV.  
Such a difference in mass scales may indicate that 
neutrino masses are of Majorana type.
Second, the WMAP data have shown that more than 
one fifth of the energy density of the Universe 
is occupied by dark matter. 
If the nature of dark matter is a weakly interacting 
massive particle (WIMP), the observed thermal relic abundance 
naturally suggests that the dark matter mass is 
around the EW scale. 
Such a mass would be generated by 
the physics just above the scale of the EWSB, the TeV scale. 

We here address the following questions:  
\begin{itemize}
\item What is the origin of the mass scale of neutrinos 
   and the WIMP dark matter? 
\item What is the relation between these scales and that of the EWSB? 
\end{itemize}
If tiny masses of the left-handed neutrinos 
are of Majorana type,
they would be generated as higher 
dimensional operators $LL\Phi\Phi/\Lambda$ at low energy 
where $L$, $\Phi$ and $\Lambda$ are 
respectively the lepton doublet field, the Higgs doublet field 
and a dimensionful parameter. 
Such operators can be realized at tree-level in three different ways; 
1) the exchange of $SU(2)_W$ singlet right-handed (RH) 
neutrinos \cite{Seesaw1}, 2) that of $SU(2)_W$ triplet scalar fields \cite{Seesaw2},
and 3) $SU(2)_W$ triplet fermions \cite{Seesaw3}.
In these tree-level mechanisms, very large masses of RH neutrinos or triplet scalars/fermions 
as compared to the scale of EWSB are required if the coupling constants are not too small.
The lowered masses are allowed in so-called radiative seesaw models, 
where the neutrino masses are generated at 
one-loop~\cite{Zee,Ma}, two-loop~\cite{ZeeBabu}, three-loop level~\cite{KNT,AKSletter}, 
or in seesaw models with  higher order (dimension $ > 5$) 
operators~\cite{de Gouvea:2007xp}.  

In a class of the radiative seesaw models~\cite{Ma,KNT,AKSletter}, a $Z_2$ parity 
is imposed to RH neutrinos to forbid the Yukawa coupling for neutrinos at tree-level.
The $Z_2$ parity also plays a role to stabilize the dark matter candidate. 
The scale of Majorana masses of the RH neutrinos is 
that of lepton number violation. 
If it comes from spontaneous breaking of 
an additional gauge symmetry such as $U(1)_{B-L}$, 
its breaking would be at the TeV scale 
in these models with TeV scale RH neutrinos. 

In this paper, we consider 
a simple scenario to explain the mass scales 
of the WIMP dark matter and neutrino simultaneously, 
in which $U(1)_{B-L}$ is spontaneously broken at the 
TeV scale by developing the vacuum expectation value (VEV) 
of the scalar field $S$. 
The $Z_2$-odd RH neutrinos then receive 
the mass $m_{N_R} \sim y_R \langle S \rangle$ 
of the electroweak scale~\cite{Khalil:2006yi,Iso:2009ss}, 
and the lightest one is the dark matter whose stability 
is guaranteed by the $Z_2$ parity. 
The tiny neutrino masses are generated 
via the one-loop induced dimension-six operator of $LL\Phi\Phi S/\Lambda^2$.
The generated mass can be 
written as 
\begin{eqnarray}
  m_{\nu_L} \sim c \left(\frac{1}{16\pi^2}\right)\left(\frac{v}{M}\right)^2 
 \langle S \rangle, 
\label{mnu2}
\end{eqnarray}
where $v \simeq 246$ GeV, $M$ represents the mass scale of the heaviest 
particle in the loop diagram and the coefficient $c$ 
is a dimensionless parameter. 
As compared to the simplest seesaw scenario, generated 
neutrino masses have the suppression factors $1/(16\pi^2)$ and  
$(m_{N_R}/M)$ (if $m_{N_R} \ll M$), so that for $M$ to be the TeV scale 
the smallness of neutrino masses can be more naturally 
explained.
Consequently, spontaneous breaking of the 
$U(1)_{B-L}$ is Mother of mass for both  
dark matter and neutrino. 

We here propose the minimal model where the scenario described 
above is realized without contradicting the current data.
The model is invariant under the gauge symmetry 
$SU(3)_C \times SU(2)_W \times U(1)_Y \times U(1)_{B-L} $.
Although the gauge group in this model is the same 
as that in Ref.~\cite{Li:2010rb}, 
we introduce neither additional fermions nor 
a nonrenormalizable term in Lagrangian.
Instead, the unbroken $Z_2$ parity is imposed 
in our model. 
In addition to the $B-L$ gauge boson $Z'$, we introduce 
the second $SU(2)_W$ scalar doublet $\eta$ which is $Z_2$ odd,  
the singlet scalar boson $S$ with a $B-L$ charge 
and three $Z_2$-odd RH neutrinos $N_R^\alpha$ ($\alpha=1$-$3$).  
The particle properties under these symmetries are summarized 
in Table.~\ref{particle}.
\begin{table}
\begin{center}
  \begin{tabular}{c|cccccc|ccc}
   \hline
   & $Q^i$ & $d_R^{i}$ & $u_R^{i}$ & $L^i$ & $e_R^i$ & 
   $\Phi$ & $\eta$ & $S$ & $N_{R}^{\alpha}$  \\\hline 
$SU(3)_C$ & $3$ & $3$ & $3$ & $1$ & $1$ & $1$ & $1$ & $1$ & $1$ \\\hline
$SU(2)_W$ & $2$ & $1$ & $1$ & $2$ & $1$ & $2$ & $2$ & $1$ & $1$ \\\hline
$U(1)_{Y}$ & $1/6$ & $-1/3$ & $+2/3$ & $1/2$ & $-1$ & $1/2$ & $1/2$ & $0$ & $0$ \\\hline
$U(1)_{B-L}$ & $1/3$ & $1/3$ & $1/3$ & $-1$ & $-1$ & $0$ & $0$ & $+2$ & $-1$ \\\hline
$Z_2$   & $+$ & $+$ & $+$ & $+$ & $+$ & $+$ & $-$ & $+$ & $-$ \\ \hline  
   \end{tabular}
\end{center}
  \caption{Particle properties. }
  \label{particle}
 \end{table}

The interaction part in the model is described as
\begin{eqnarray}
\mathcal{L}_{\mathrm{int}} = \mathcal{L}^{\rm SM}_{\mathrm{Yukawa}} + 
\mathcal{L}_{N} - V(\Phi, \eta, S),
\end{eqnarray}
where $\mathcal{L}^{\rm SM}_{\mathrm{Yukawa}}$ is the SM Yukawa interaction, and 
\begin{eqnarray}
 {\cal L}_{N} = \sum_{\alpha=1}^3 \left(\sum_{i=1}^3  
   g_{i\alpha} \overline{L}^i \tilde{\eta} N_R^\alpha  
   - \frac{y_R^\alpha}{2}  
     \overline{N}_R^\alpha S N_R^\alpha + {\rm h.c.}\right),
\end{eqnarray}
 with $\tilde{\eta}=i\tau_2 \eta^\ast$.  
 Without loss of generality, $y_R$ can be taken to be flavor diagonal. 
 Under the $Z_2$ parity, neutrino Yukawa couplings 
 among $L, \Phi$ and $N_R^\alpha$ are forbidden. 
 The scalar potential $V(\Phi, \eta, S)$ is given by
\begin{eqnarray} 
&& V(\Phi, \eta, S) = + \mu_1^2 |\Phi|^2 + \mu_2^2 |\eta|^2 + \mu_S^2 |S|^2 
+\lambda_1 |\Phi|^4  \nonumber\\
&&
  + \lambda_2|\eta|^4 + \lambda_3 |\Phi|^2|\eta|^2 + \lambda_4|\Phi^{\dagger} \eta|^2 
  + \frac{\lambda_5}{2} \left[ (\Phi^{\dagger} \eta)^2 + {\rm h.c.} \right] \nonumber \\
  && +\lambda_S |S|^4 + \tilde{\lambda} |\Phi|^2|S|^2 
     + \lambda |\eta|^2|S|^2,
\label{HiggsPotential} 
\end{eqnarray}
where $\lambda_5$ can be taken as real. We assume that 
$\mu_1^2$ and $\mu_S^2$ are negative while $\mu_2^2$ is positive.

The $U(1)_{B-L}$ symmetry is spontaneously broken when 
the $S$ develops the VEV $\langle S\rangle = v_S/\sqrt{2}$ 
at the TeV scale~\cite{Khalil:2006yi,Iso:2009ss}.
We assume for simplicity that kinetic mixing  \cite{Holdom} between $U(1)_Y$ and 
$U(1)_{B-L}$ gauge bosons is small such that it satisfies the EW precision measurement.
The $Z'$ boson then acquires the mass
 $M_{Z'}^2 = 4 g_{B-L}^2 v_S^2$ where $g_{B-L}^{}$ is the 
gauge coupling constant for $U(1)_{B-L}$.
From the LEP experiment, the lower bound 
on the $Z'$ boson mass has been found to 
be
$v_S \gtrsim 3$-$3.5$  {\rm TeV}~\cite{LEPbound1,LEPbound2}. 
Recent bound from Tevatron is comparable to the LEP 
bound~\cite{Basso:2008iv}.
The $Z_2$-odd RH neutrinos 
$N_R^{\alpha}$ also obtain masses from the $U(1)_{B-L}$ breaking as    
\begin{equation}
 m_{N_R^\alpha} = -y_{R}^\alpha \frac{v_S}{\sqrt{2}} .
\end{equation}
After the EWSB, 
the neutral components of the $Z_2$ even scalar fields 
are parameterized as
\begin{eqnarray} 
 \Phi^0 =  \frac{1}{\sqrt{2}}(v + \phi + i z) ,  \hspace{4mm} 
  S      =  \frac{1}{\sqrt{2}} (v_S^{} + \phi_S^{} + i z_S^{}), 
\end{eqnarray}
 where $z$ and $z_S^{}$ are the Nambu-Goldstone bosons absorbed by  
  the longitudinal modes of the weak and $U(1)_{B-L}$ gauge bosons 
 $Z$ and $Z'$, respectively.  
We define the ratio of the two VEVs as 
  $\tan\beta = v_S^{}/v$.
The mass matrix of $\phi$ and $\phi_S^{}$ is 
 diagonalized by a mixing angle $\alpha$;  
\begin{eqnarray}
 \begin{pmatrix}
   h \\
   H
 \end{pmatrix}
 =
 \begin{pmatrix}
  \cos\alpha && -\sin\alpha \\
  \sin\alpha && \cos\alpha
 \end{pmatrix}
 \begin{pmatrix}
  \phi \\
  \phi_S^{}
 \end{pmatrix},
\end{eqnarray}
where $h$ and $H$ represent the eigenstates corresponding to 
the mass eigenvalues 
\begin{eqnarray}
 m_{h}^2 &=& 2 ( \lambda_1 c_\alpha^2 + \lambda_S s_\alpha^2 \tan^2\beta  
-  \tilde{\lambda} s_\alpha c_\alpha \tan\beta) v^2, \\
 m_{H}^2 &=& 2 ( \lambda_1 s_\alpha^2  + \lambda_S c_\alpha^2 \tan^2\beta 
 + \tilde{\lambda} s_\alpha c_\alpha \tan\beta) v^2, 
\end{eqnarray}
where $c_\alpha=\cos\alpha$ and $s_\alpha=\sin\alpha$. 
The mixing angle $\alpha$ is a free parameter. 
The constraints on the Higgs mixing from the precision measurements 
have been derived in Ref.~\cite{Dawson:2009yx}.
The Yukawa interactions for $h$ and $H$ with $N_R^\alpha$ are then given by 
\begin{eqnarray}
\mathcal{L}_{\mathrm{yukawa}} =  
  - \frac{1}{2}  y_R^\alpha
     \bar{N}_R^\alpha \frac{(-h\sin\alpha  + H\cos\alpha)}{\sqrt{2}} N_R^\alpha 
+ \cdot\cdot\cdot .
\end{eqnarray}
The component fields of the $Z_2$-odd isospin doublet $\eta$ 
are parametrized as
\begin{equation}
 \eta = \left(
\begin{array}{c}
  H'^+ \\
  \frac{1}{\sqrt{2}}(H'+ i A')
\end{array}
 \right),  
\end{equation}
whose masses are mainly determined by the invariant mass 
parameter $\mu_2$ as long as $\mu_2^2 \gg \lambda_i v^2 \sim \lambda v_S^2$.
\begin{figure}[t]
\begin{center}
  \epsfig{file=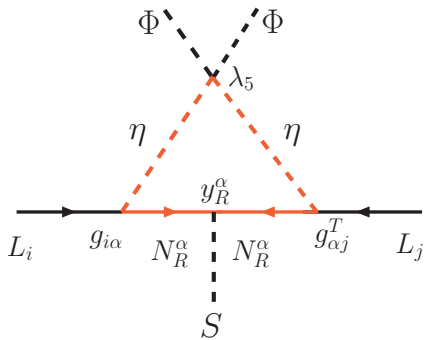,width=6cm}
\end{center}
  \caption{The one-loop diagram of the dimension-six operator which 
generates neutrino masses.}
  \label{diagram}
\end{figure}


The tiny neutrino masses are generated 
via the one-loop induced dimension-six operator 
$L L S \Phi \Phi$ shown in 
Fig.~{\ref{diagram}}.
The induced mass matrix is evaluated as 
\begin{equation}
 m_{\nu_L}^{ij} \simeq  
  \frac{\lambda_5}{8 \pi^2} 
  \left( \sum_{\alpha=1}^3 g_{i \alpha} y_R^\alpha g_{\alpha j}^T \right) 
  \left(\frac{v}{m_{\phi'}}\right)^2  v_S, 
\label{NuMassMatrix}
\end{equation}
 for $m_{\phi'}^2 \gg m_{N_R^\alpha}^2$, 
where $m_{\phi'}$ denotes the scale of the quasi degenerate masses of $H'$ and $A'$.
For $v/m_{\phi'} \sim 10^{-1}$ and 
$\lambda_5 \sim g_{i\alpha} \sim y_R^\alpha \sim 10^{-2}$, 
the correct mass scale ($\sim 0.1$ eV) can be realized from 
the TeV scale $v_S$.  
It is easy to see that the observed neutrino oscillation data can be 
reproduced by the result in Eq.~(\ref{NuMassMatrix}),
because the flavor structure is the 
same as that of the standard seesaw mechanism\cite{Perez:2009mu}.

\begin{figure}[t]
\begin{center}
  \epsfig{file=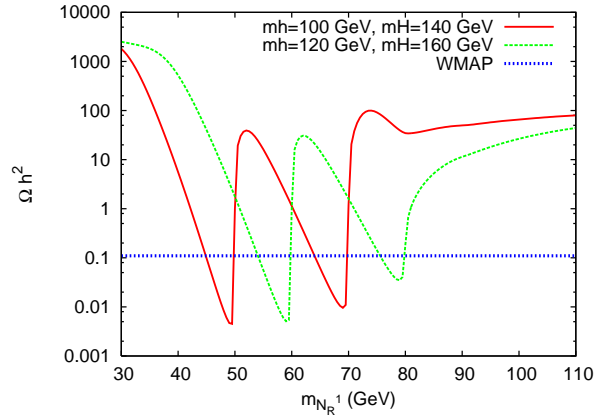,width=8cm,angle=+0}
\end{center}
  \caption{The abundance of $N_R^1$ as a function of 
   the mass $m_{N_R^1}$. The masses of $h$ and $H$ are taken 
   to be $m_h=100$ GeV and $m_H=140$ GeV (red curve), and 
         $m_h=120$ GeV and $m_H=160$ GeV (green curve) for 
 $\sin\alpha=1/\sqrt{2}$ and $\tan\beta=15$.}
  \label{fig:omegah2}
\end{figure}

The dark matter candidates are the lightest RH neutrino $N_{R}^1$
and the lightest neutral component of the $Z_2$-odd  doublet ($H'$ or $A'$). 
In the model by Ma~\cite{Ma},   
the thermal relic density has been 
estimated for $H' (A')$~\cite{Hambye:2006zn} 
as well as $N_{R}^1$~\cite{Kubo:2006yx}. 
However, the scenario for the $N_{R}^1$ dark matter suffers from 
the constraint from lepton flavor violation (LFV) such as 
$\mu \rightarrow e\gamma$.  
The sufficient annihilation of dark matter requires 
large couplings $g_{i\alpha}$ whereas the null result for LFV 
gives the upper bound on these couplings~\cite{Kubo:2006yx}. 
To avoid the difficulty, additional mechanism 
such as co-annihilation has to be introduced~\cite{Suematsu}. 
On the contrary, 
in our model with the spontaneously broken $U(1)_{B-L}$ symmetry, 
the neutral component of $S$ 
can largely mix with the SM Higgs boson $\phi$. 
Consequently,  $N_{R}^1$ can annihilate via 
$N_R^1N_R^1 \to h (H) \to f\overline{f}$~\cite{OkadaSeto:B-L}, 
so that the required relic abundance $\Omega_{N_{R}^1}h^2 \simeq 0.11$ 
can easily be attained when $m_{N_{R}^1} \sim m_{h}/2$ or $m_{H}/2$ 
without contradicting the LFV constraints.
Therefore, 
$N_R^1$ can be the dark matter.  
In Fig.~{\ref{fig:omegah2}}, the thermal relic abundance of 
$N_R^1$ is shown as a function of $m_{N_R}$ for 
$m_h=100$ ($120$) GeV, $m_H=140$ ($160$) GeV, 
$\sin\alpha=1/\sqrt{2}$ and $\tan\beta=15$.  
It is seen that the relic abundance of $N_R^1$ is significantly reduced 
around $m_{N_R} = 50~(60)$ and $70~(80)$ GeV, respectively, and becomes consistent 
with the observed abundance of the dark matter. The RH neutrinos annihilate 
mainly via the S-channel exchange of the Higgs scalars. The annihilation of the RH neutrinos into 
the SM particles is resonantly enhanced through S-channel exchange of $h$ or $H$ 
when $m_h \simeq 2 m_{N_R}$ and $m_h \simeq 2 m_{N_R}$. Note that RH neutrinos can also 
annihilate via the $B-L$ gauge boson exchange. However this process is less important 
because the cross section of the $Z'$ exchange, $\langle \sigma v\rangle$, is 
proportional to $1/v_S^4$ and is much smaller than that of the Higgs exchange.


As a successful scenario, 
we consider the following parameters. 
First, $v_S$ 
is supposed to be about $3.7$ TeV. 
Second, $m_h$, $m_H$ and $\alpha$ are 
taken as $100$ GeV, $120$ GeV and $\pi/4$, respectively. Third, the $Z'$ mass is assumed to be 
$0.5$-$1$ TeV, and the masses of $H'$, $A'$ and $H'^+$ are commonly taken to be a few TeV. 
Finally, $m_{N_R^1}$ and $m_{N_R^{2,3}}$ are taken as $46$ GeV and a few hundred GeV, respectively.

Phenomenological predictions are discussed  
in order: \\
%
%
\hspace*{6pt}
I) A characteristic feature of the Higgs bosons $h$ and $H$ 
   with large mixing is that all the 
   coupling constants of $h$ ($H$) to the SM particles 
   are multiplied by $\sin\alpha$  ($\cos\alpha$) 
   as compared to the SM ones.  
   This is similar to the Type-I two Higgs doublet model (THDM) 
   without charged Higgs bosons.  
   For maximal mixing $\sin\alpha \sim 1/\sqrt{2}$, 
   the visible widths of $h$ and $H$ are about a half of that for the 
   SM Higgs boson. 
   The decay pattern can be discriminated from the other 
   types of the Yukawa coupling in the THDM\cite{4Yukawa}.
   Production rates of $h$ ($H$) at the Tevatron and the LHC are 
   about 50\% smaller than the SM value.\\
%
%
\begin{figure}[t]
\begin{center}
\begin{tabular}{c}
 \includegraphics[height=55mm]{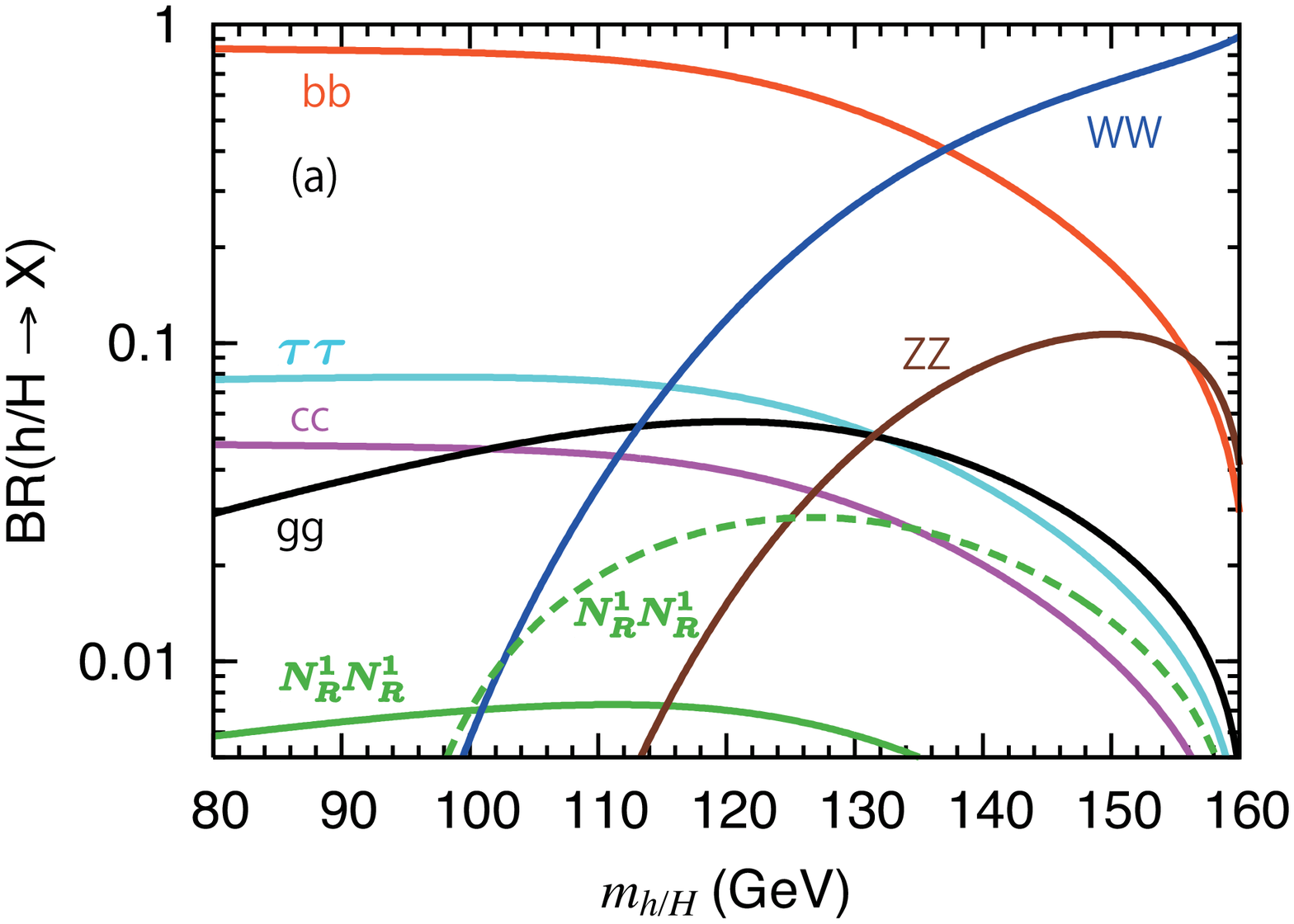} \\
 \includegraphics[height=55mm]{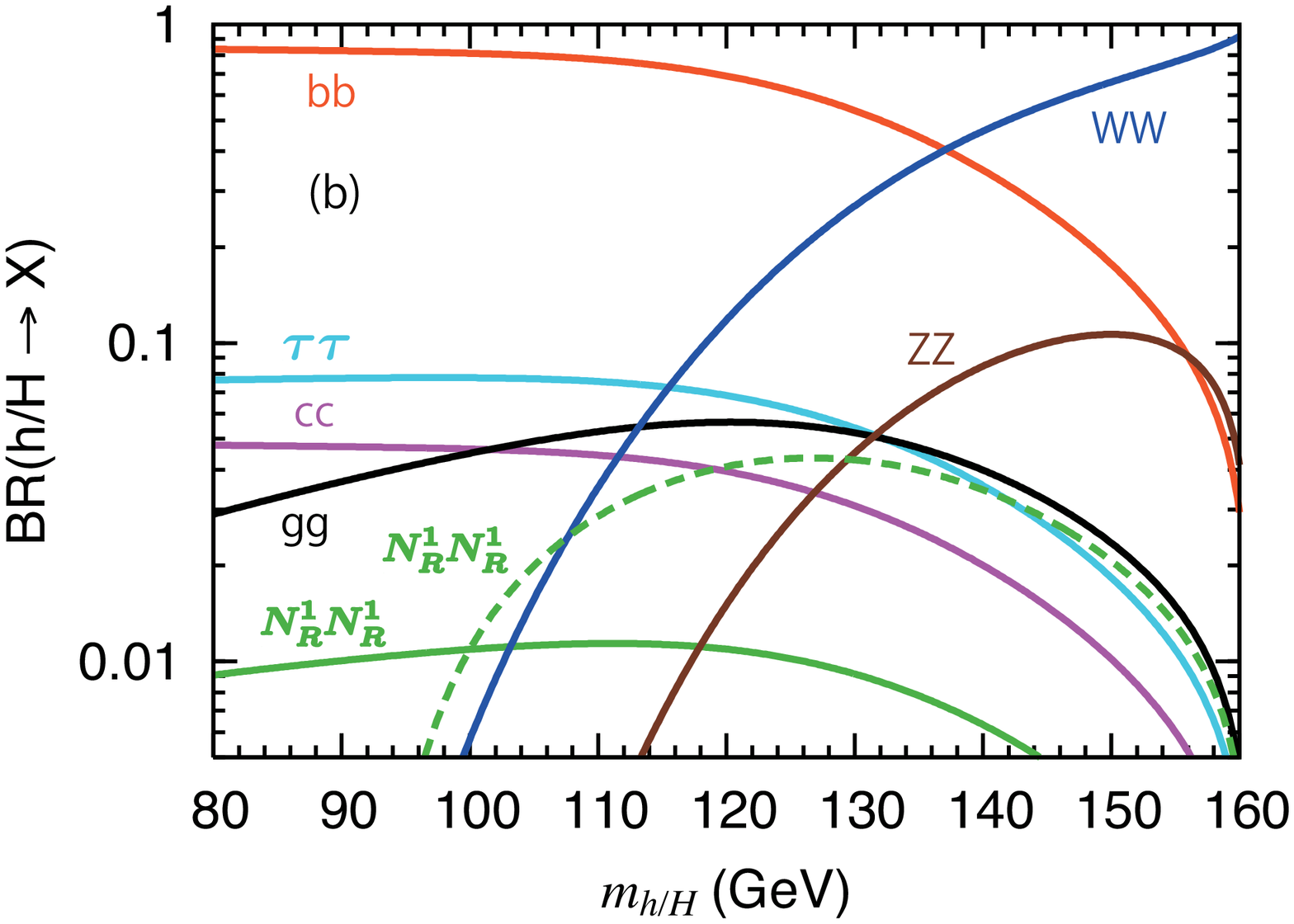}
\end{tabular}
\caption{Branching rations of $h$ and $H$ as a function of their masses. 
$\tan\beta$ is taken to be $15$ in the figures (a) and $12$ in (b), respectively. 
The solid curve for $N_R^1 N_R^1$ represents the branching ratio of $h$, while 
the dashed one does that of $H$. The mass of the lightest RH neutrino is fixed as the 
half of the mass of $h$ for the decay of $h$, while it is fixed as $46$ GeV for that of $H$.}
\label{fig:higgs-branching}
\end{center}
\end{figure}
\hspace*{6pt}
II) The Higgs bosons $h$ and $H$ also couple to the dark matter $N_R^1$, 
    so that they decay into a dark matter pair if kinematically 
    allowed\cite{Bento:2000ah}. In Fig.~\ref{fig:higgs-branching}, the branching ratios 
    of the decays of $h$ and $H$ are shown as a function of $m_h$ and $m_H$. 
    $\tan\beta = 15$ and $12$ is taken in Fig.~3(a) and 3(b), respectively. 
    The solid curve for $N_R^1 N_R^1$ represents 
    the decay branching ratio of $h$, while the dashed one does that of $H$. 
    As shown in Fig.~\ref{fig:omegah2}, the relic abundance of $N_R^1$ can be consistent 
    with that of the dark matter when its mass is slightly smaller or larger than the half of the mass 
    of $h$ or $H$. We assume here that the mass of $N_R^1$ is slightly smaller than the half of the 
    mass of $h$. The mass of $N_R^1$ is fixed as the half of the mass of $h$ minus $4$ 
    GeV for the decay of $h$ while it is fixed as $46$ GeV for that of $H$. 
    The invisible decay of $h$ and $H$ can reach to 0.7 \% (2.8 \%) for $h~(H)$ 
    in Fig. 3(a), and 1.1 \% (4.3 \%) in Fig. 3(b).
    For $v_S \sim 3.7$ TeV, which correspond to $\tan\beta = 15$, the coupling constant 
    $y_R$ is determined as about 0.01 
    to obtain the correct mass $m_{N_R^1} \sim m_h/2$ or $m_H/2$.
    For $m_{N_R^1} =46$ GeV, the invisible decay $h (H) \to N_R^1 N_R^1$ 
    is then evaluated as 0.7 \% (2.6 \%). 
    The invisible decay of the SM-like Higgs boson can be 
    detected if it is larger than 25 \% at the LHC~\cite{LHC} and a few \% 
    at the International Linear Collider (ILC)~\cite{ILC}, respectively. 
    Feasibility of dark matter at colliders as well as 
    direct searches  is discussed in terms of the 
    simple Higgs portal 
    dark matter model in Ref.~\cite{higgsportal}. 
   Studies on the RH neutrino production in the model
   without the $Z_2$ parity are seen in Ref.~\cite{Basso:2008iv,Huitu:2008gf}.
   In our model, $N_{R}^1$ is dark matter, 
   while heavier RH neutrinos $N_{R}^2$ and $N_{R}^3$ can be tested 
   via the decay into $N_{R}^1$ with a lepton pair.\\
%
%
\hspace*{6pt}
III) The existence of the $Z'$ boson is   
     another difference from the model by Ma~\cite{Ma}. 
     Its production at the LHC in the minimal $U(1)_{B-L}$ model 
     has been discussed~\cite{Emam:2007dy,Basso:2008iv}. 
     If $Z'$ is lighter than a few TeV, it would be detected 
     at the LHC. The production cross section of 
    $Z'$ can be ${\cal O}(1)$ pb for $m_{Z'}^{} \sim 1$~TeV.
    Decays of $Z'$ into SM particles 
    are proportional to the $B-L$ charges.   
    The branching ratios of 
    $Z' \to q \bar q$, $\ell^+\ell^-$, $\nu_L \bar \nu_L$, 
    $N_R N_R$ and $\phi_S \phi_S$ are given approximately by 
    0.2, 0.3, 0.15, 0.15 and 0.2, respectively.  
    The branching ratio of $N_R^{2,3} \to N_R^{1} \nu_L \nu_L$ 
    is about 0.5.   
    The invisible decay of $Z'$ is then 0.225. 
    This is a definite prediction in our model. 
    It is expected that the invisible decay as well as characteristic 
    branching ratios of visible decays can be tested 
    at the LHC and the ILC.   \\
%
%
\hspace*{6pt}
IV) An important constraint on radiative neutrino 
mass models comes from LFV processes  
such as $\mu \to e\gamma$.
Since the typical scale of $g_{i\alpha}$ 
is of order of $10^{-2}$, 
the present bound can be satisfied,  but 
this would be testable by the new data 
from flavor experiments 
such as MEG. 
\\
%
\hspace*{6pt}
V) Phenomenology of the $Z_2$-odd scalar bosons 
$H'$, $A'$, $H'^\pm$ has been studied in various models~\cite{inert}.  
In our present scenario, their masses are at the TeV scale.
Although their direct detection would be difficult at the LHC,  
their indirect effects appear in the three-body decay of heavier 
right-handed neutrinos $N_R^{2(3)} \rightarrow N_R^1 \ell^-\ell^+$ 
that can be tested at colliders.\\
\hspace*{6pt}
Therefore, the model can be tested and discriminated 
from the other radiative seesaw models, 
the minimal $U(1)_{B-L}$ model, and the 
other extended Higgs models such as THDMs. 

Finally, we give a comment on the possibility of baryogenesis.
In this minimal model, the $Z_2$-even part in the Higgs sector
is composed of the isospin doublet $\Phi$ and the singlet $S$, 
so that there is no additional CP phase.  
In addition, coupling constants in the Higgs sector are so small 
that first order phase transition cannot realize. 
Thus, for successful electroweak baryogenesis\cite{ewbg1}, 
the model has to be extended, for example, 
by introducing additional $Z_2$-even scalar doublets\cite{ewbg2,AKSletter}.  
On the other hand, TeV scale leptogenesis 
may be an alternative way.  Various 
mechanisms have been proposed such as resonant 
leptogenesis~\cite{ResonantLeptogenesis}
or three-body decay~\cite{3DecayLeptogenesis}.  
In these scenarios, however, considerable fine-tuning 
would be required.

We have discussed the minimal model with the mass generation 
mechanism for dark matter and neutrino based on  
the $U(1)_{B-L}$ symmetry and the $Z_2$ parity. 
Spontaneous $U(1)_{B-L}$ breaking at the TeV scale gives 
the Majorana masses of $N_R^\alpha$, and the lightest 
one can be a WIMP dark matter.
Its thermal relic abundance explains the WMAP result.  
Tiny neutrino masses are radiatively generated 
via the dimension-six operator without excessive fine tuning. 
The model can be tested at future experiments.
The detailed analyses are 
shown elsewhere~\cite{kss_full}.


This work was supported in part by Grant-in-Aid for Scientific Research, 
Japan Society for the Promotion of Science, Nos. 22244031 and 19540277~(SK), 
and by the scientific research grants from Hokkai-Gakuen~(OS).
TS is partially supported by Yukawa Memorial Foundation.



\begin{thebibliography}{1}

\bibitem{Seesaw1}
  T.~Yanagida,
   in \textit{Proceedings of Workshop on the Unified Theory and
   the Baryon Number in the Universe}, Tsukuba, Japan,
   edited by A.~Sawada and A.~Sugamoto (KEK, Tsukuba, 1979), p 95;
  M.~Gell-Mann, P.~Ramond, and R.~Slansky,
   in \textit{Supergravity},
   Proceedings of Workshop, Stony Brook, New York, 1979, edited by
   P.~Van~Nieuwenhuizen and D.~Z.~Freedman
   (North-Holland, Amsterdam, 1979), p 315;
  R.~N.~Mohapatra and G.~Senjanovic, Phys. Rev. Lett. {\bf 44}, 912 (1980).

\bibitem{Seesaw2}
   W.~Konetschny and W.~Kummer,
   Phys.\ Lett.\  B {\bf 70}, 433 (1977).
   T.~P.~Cheng and L.~F.~Li,
   Phys.\ Rev.\  D {\bf 22}, 2860 (1980).
   J.~Schechter and J.~W.~F.~Valle,
  Phys.\ Rev.\  D {\bf 22}, 2227 (1980).
  
\bibitem{Seesaw3}
  R.~Foot, H.~Lew, X.~G.~He and G.~C.~Joshi,
  Z.\ Phys.\  C {\bf 44}, 441 (1989).


\bibitem{Zee}
  A.~Zee,
  Phys.\ Lett.\  B {\bf 93}, 389 (1980)
  [Erratum-ibid.\  B {\bf 95}, 461 (1980)].
  
  
\bibitem{Ma}
  E.~Ma,
  Phys.\ Rev.\  D {\bf 73}, 077301 (2006).


\bibitem{ZeeBabu}
   A.~Zee, Nucl.\ Phys.\ B {\bf 264}, 99 (1986).
   K.~S.~Babu, Phys.\ Lett.\ B {\bf 203}, 132 (1988).

  

\bibitem{KNT}
  L.~M.~Krauss, S.~Nasri and M.~Trodden,
  Phys.\ Rev.\  D {\bf 67}, 085002 (2003);  
  K.~Cheung and O.~Seto,
  Phys.\ Rev.\  D {\bf 69}, 113009 (2004).

\bibitem{AKSletter}
  M.~Aoki, S.~Kanemura and O.~Seto,
  Phys.\ Rev.\ Lett.\  {\bf 102}, 051805 (2009); 
  Phys.\ Rev.\  D {\bf 80}, 033007 (2009).


\bibitem{de Gouvea:2007xp}
  A.~de Gouvea and J.~Jenkins,
  Phys.\ Rev.\  D {\bf 77}, 013008 (2008); 
  F.~Bonnet, D.~Hernandez, T.~Ota and W.~Winter,
  JHEP {\bf 0910}, 076 (2009); 
  S.~Kanemura and T.~Ota,
  Phys.\ Lett.\  B {\bf 694}, 233 (2010).

\bibitem{Khalil:2006yi}
  S.~Khalil,
  J.\ Phys.\ G {\bf 35}, 055001 (2008).

\bibitem{Iso:2009ss}
  S.~Iso, N.~Okada and Y.~Orikasa,
  Phys.\ Lett.\  B {\bf 676}, 81 (2009); 
  Phys.\ Rev.\  D {\bf 80}, 115007 (2009). 
  
\bibitem{Holdom}
  B.~Holdom,
  Phys.\ Lett.\  B {\bf 166}, 196 (1986).

\bibitem{Li:2010rb}
  T.~Li and W.~Chao,
  Nucl.\ Phys.\  B {\bf 843}, 396 (2011).

\bibitem{LEPbound1}
  M.~S.~Carena, A.~Daleo, B.~A.~Dobrescu and T.~M.~P.~Tait,
  Phys.\ Rev.\  D {\bf 70}, 093009 (2004).

\bibitem{LEPbound2}
  G.~Cacciapaglia, C.~Csaki, G.~Marandella and A.~Strumia,
  Phys.\ Rev.\  D {\bf 74}, 033011 (2006). 

\bibitem{Basso:2008iv}
  L.~Basso, A.~Belyaev, S.~Moretti and C.~H.~Shepherd-Themistocleous,
  Phys.\ Rev.\  D {\bf 80}, 055030 (2009).

\bibitem{Dawson:2009yx}
  S.~Dawson and W.~Yan,
  Phys.\ Rev.\  D {\bf 79}, 095002 (2009).

\bibitem{Perez:2009mu}
  P.~F.~Perez, T.~Han and T.~Li,
  Phys.\ Rev.\  D {\bf 80}, 073015 (2009).


\bibitem{Hambye:2006zn}
  T.~Hambye, K.~Kannike, E.~Ma and M.~Raidal,
  Phys.\ Rev.\  D {\bf 75}, 095003 (2007).

\bibitem{Kubo:2006yx}
  J.~Kubo, E.~Ma and D.~Suematsu,
  Phys.\ Lett.\  B {\bf 642}, 18 (2006).

\bibitem{Suematsu}
  D.~Suematsu, T.~Toma and T.~Yoshida,
  Phys.\ Rev.\  D {\bf 79}, 093004 (2009).


\bibitem{OkadaSeto:B-L}
  N.~Okada and O.~Seto,
  Phys.\ Rev.\  D {\bf 82}, 023507 (2010).

\bibitem{4Yukawa} 
 V.~D.~Barger, J.~L.~Hewett and R.~J.~N.~Phillips, Phys.\ Rev.\ D {\bf 41}, 3421 (1990);
 Y.~Grossman, Nucl.\ Phys.\ B {\bf 426}, 355 (1994);
  M.~Aoki, S.~Kanemura, K.~Tsumura and K.~Yagyu,
  Phys.\ Rev.\  D {\bf 80}, 015017 (2009).

\bibitem{Bento:2000ah}  
  M.~C.~Bento, O.~Bertolami, R.~Rosenfeld and L.~Teodoro,
  Phys.\ Rev.\  D {\bf 62}, 041302 (2000).

\bibitem{LHC}
  B.~Di~Girolamo and L.~Neukermans, Atlas Note ATL-PHYS-2003-006;
  M.~Warsinsky [ATLAS Collaoration], J. Phys. Conf. Ser. 110, 072046 (2008).

\bibitem{ILC}
 M.~Schumacher, Report No. LC-PHSM-2003-096.
 
\bibitem{higgsportal}

  X.~G.~He, et al., 
  Phys.\ Lett.\  B {\bf 688}, 332 (2010);   
  M.~Farina, D.~Pappadopulo and A.~Strumia,
  Phys.\ Lett.\  B {\bf 688}, 329 (2010); 
  K.~Cheung and T.~C.~Yuan,
  Phys.\ Lett.\  B {\bf 685}, 182 (2010);
%
  M.~Aoki, S.~Kanemura and O.~Seto,
  Phys.\ Lett.\ B\ {\bf 685}, 313 (2010);
%
  S.~Kanemura, S.~Matsumoto, T.~Nabeshima and N.~Okada,
  Phys.\ Rev.\  D {\bf 82}, 055026 (2010).

\bibitem{Huitu:2008gf}
  K.~Huitu, S.~Khalil, H.~Okada and S.~K.~Rai,
  Phys.\ Rev.\ Lett.\  {\bf 101}, 181802 (2008).

\bibitem{Emam:2007dy}
  W.~Emam and S.~Khalil,
  Eur.\ Phys.\ J.\  C {\bf 522}, 625 (2007).

\bibitem{inert}
  R.~Barbieri, L.~J.~Hall and V.~S.~Rychkov,
  Phys.\ Rev.\  D {\bf 74}, 015007 (2006); 
  Q.~H.~Cao, E.~Ma and G.~Rajasekaran,
  Phys.\ Rev.\  D {\bf 76}, 095011 (2007).

\bibitem{ewbg1}

e.g., A.~G.~Cohen, D.~B.~Kaplan and A.~E.~Nelson,
Ann.\ Rev.\ Nucl.\ Part.\ Sci.\  {\bf 43}  27 (1993);~
%
M.~Quiros,
Helv.\ Phys.\ Acta {\bf 67} 451 (1994);~
%
V.~A.~Rubakov and M.~E.~Shaposhnikov,
Usp.\ Fiz.\ Nauk {\bf 166} 493 (1996)
[Phys.\ Usp.\  {\bf 39} 461 (1996)].


\bibitem{ewbg2}

  J.~M.~Cline, K.~Kainulainen and A.~P.~Vischer,
  Phys.\ Rev.\  D {\bf 54}, 2451 (1996);
  S.~Kanemura, Y.~Okada and E.~Senaha,
  Phys.\ Lett.\  B {\bf 606}, 361 (2005);
%
   L.~Fromme, S.~J.~Huber and M.~Seniuch,
  JHEP {\bf 0611}, 038 (2006).

\bibitem{ResonantLeptogenesis}
  A.~Pilaftsis and T.~E.~J.~Underwood,
  Nucl.\ Phys.\  B {\bf 692}, 303 (2004).

\bibitem{3DecayLeptogenesis}
  e.g., T.~Hambye,
  Nucl.\ Phys.\  B {\bf 633}, 171 (2002).

\bibitem{kss_full}
 S.~Kanemura, O.~Seto, and T.~Shimomura, in preparation.


\end{thebibliography}
\end{document}